\documentclass[sigconf]{acmart}

\AtBeginDocument{%
  \providecommand\BibTeX{{%
    \normalfont B\kern-0.5em{\scshape i\kern-0.25em b}\kern-0.8em\TeX}}}

\setcopyright{ccbysa}
\copyrightyear{2022}
\acmYear{2022}
\acmDOI{}

\acmConference[IMI Workshop'22]{Intelligent Music Interfaces: When Interactive Assistance and Augmentation Meet Musical Instruments}{April 30,
  2022}{New Orleans, LA}
\acmBooktitle{Intelligent Music Interfaces: When Interactive Assistance and Augmentation Meet Musical Instruments, April 20, 2022,New Orleans, LA}
\acmISBN{}

\usepackage{subcaption}

\usepackage{tabularx}
\usepackage{dcolumn} 
\newcolumntype{d}[1]{D{.}{.}{#1}}

\begin{document}

\title{A Vision of the Intelligent Piano of the Future}
\title{The Next Generation of Pianos}
\title{The Future of Pianos}
\title{Making Pianos Intelligent}
\title{The Supportive Piano of the Future}
\title{The Intelligent Piano of the Future}
\title{The Human-Centered Piano of the Future}
\title{The Vision of a Human-Centered Piano of the Future}
\title{The Vision of a Human-Centered Piano}


\settopmatter{authorsperrow=4}
\author{Jordan Aiko Deja}
\orcid{0001-9341-6088}
\affiliation{%
  \institution{University of Primorska}
  \city{Koper}
  \country{Slovenia}
 \postcode{6000}}
 \affiliation{%
  \institution{De La Salle University}
  \city{Manila}
  \country{Philippines}}
\email{jordan.deja@famnit.upr.si}

\author{Sven Mayer}
\affiliation{%
  \institution{LMU Munich}
 \city{Munich}
  \country{Germany}}
\email{info@sven-mayer.com}

\author{Klen Čopič Pucihar}
\affiliation{%
  \institution{University of Primorska}
  \city{Koper}
  \country{Slovenia}
  \postcode{6000}}
\email{klen.copic@famnit.upr.si}

\author{Matjaž Kljun}
\affiliation{%
  \institution{University of Primorska}
  \city{Koper}
  \country{Slovenia}
  \postcode{6000}}
\email{matjaz.kljun@famnit.upr.si}

\renewcommand{\shortauthors}{Deja et al.}

\begin{abstract}
For around 300 years, humans have been learning to play the modern piano either with a teacher or on their own. In recent years teaching and learning have been enhanced using augmented technologies that support novices. Other technologies have also tried to improve different use cases with the piano, such as composing and performing. Researchers and practitioners have showcased several forms of augmentation, from hardware improvements, sound quality, rendering projected visualizations to gesture-based and immersive technologies. Today, the landscape of piano augmentations is very diverse, and it is unclear how to describe the ideal piano and its features. In this work, we discuss how the human-centered piano -- the piano that has been designed with humans in the center of the design process and that effectively supports tasks performed on it -- can support pianists. In detail, we present the three tasks of learning, composing, and improvising in which a human-centered piano would be beneficial for the pianist.
\end{abstract}

\ccsdesc[500]{Human-centered computing~Human computer interaction (HCI)}
\ccsdesc[500]{Human-centered computing~Interaction devices}
\ccsdesc[300]{Applied computing~Sound and music computing}
\ccsdesc[300]{Applied computing~Interactive learning environments}

\keywords{piano, human-centered design, intelligent music interface, music interface}


\maketitle

\section{Introduction}
Bartolomeo Cristofori designed the first piano in the 1700s deriving it from the clavichord and harpsichords~\cite{pollens1995early}. The piano generates sounds using hammers (controlled by a keyboard) striking strings. The initial piano implementation has been evaluated and reviewed to address various issues or improve the experience in general. For instance, dimensions and measurements were standardized for supporting proper posture while playing the piano. However, further research has observed that these measurements may not be as inclusive and may be gender-biased. To address this, physical augmentations have been introduced instead to accommodate users with specific needs (e.g., adding a seat riser for shorter people, adding a cushion for comfort during prolonged usage). Other physical appendages and digital hardware (e.g. magnetic resonators, actuators~\cite{mcpherson2010magnetic}, string actuation for key sensing~\cite{mcpherson2010augmenting}, sensors for pitch control~\cite{mcpherson2013space}) have also been introduced to improve sound quality or introduce a different listening experience. Beyond the listening experience, augmentations have also been done to serve disparate set of use-cases in the piano. They come in either digital, electronic or acoustic augmentation. 

\begin{figure}[t]
\centering
  \includegraphics[width=\linewidth]{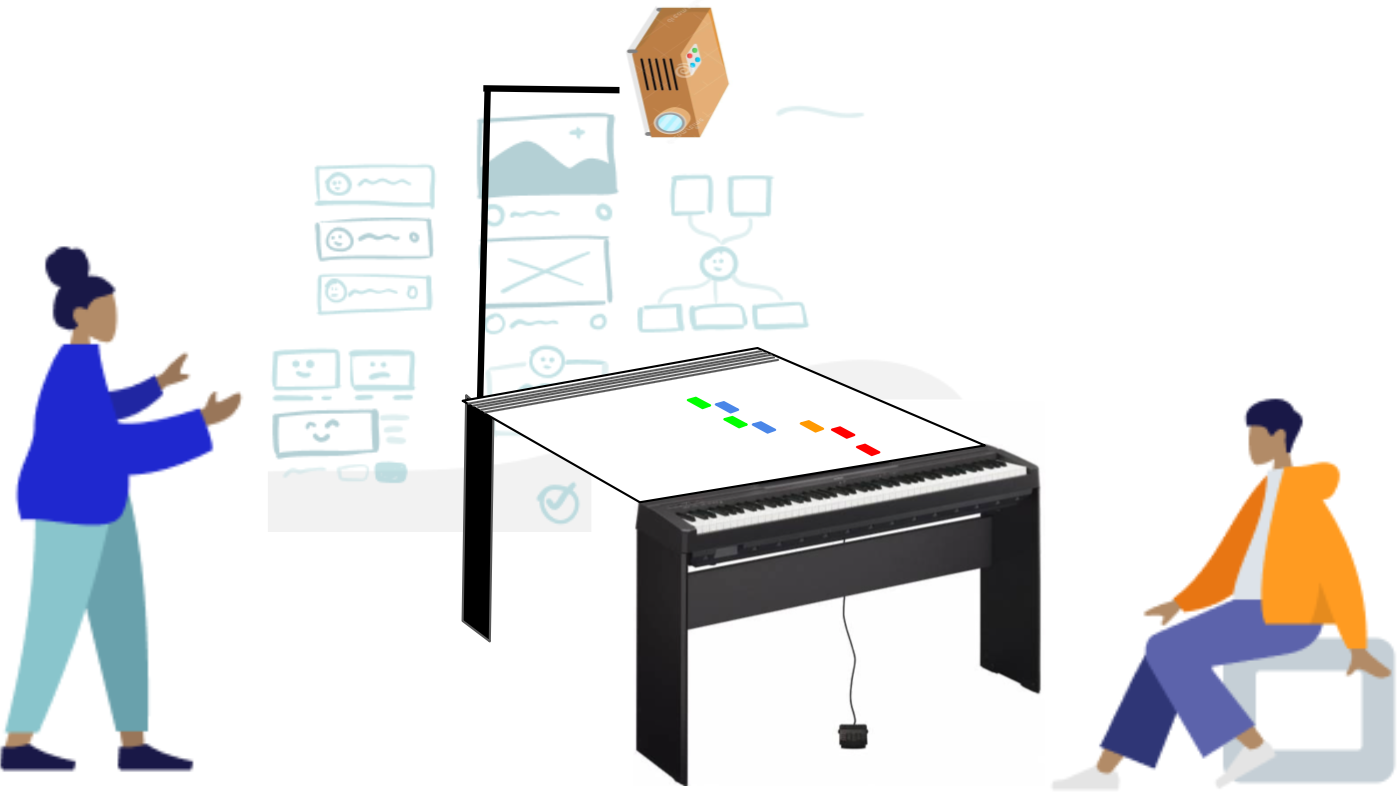}
  \caption{A human-centered piano intelligently understands the contexts behind each task of a pianist: learner, composer, performer. }
  \Description{A human-centered piano intelligently understands the contexts behind each task of a pianist: learner, composer, performer. }
  \label{fig:teaser}
\end{figure}

Generally, an augmentation is initially introduced to the piano to serve a specific purpose. For instance, electronic augmentations (e.g., audio jack or a MIDI interface) have been added to allow the piano to be listened to by large crowd or by oneself. For example, to accommodate large audiences during events such as recitals, concerts, or orchestras, the piano has to be connected to a sound system. If a pianist wants to listen to their work privately, they can listen to a recording of their own practice on the headphones. Recording, playing back music and post-processing with filters became possible too with these augmentations. At present, these augmentations appear minor or standard since they are now part of consumer-based electronic pianos.

With recent technological innovations such as the internet, artificial intelligence, and mixed reality, designers have further reshaped the piano and music listening experience~\cite{deja2020sound}. Just within the recent two decades, we have seen a wave of digital piano augmentations that cater to different use-cases and tasks: (i) vision-based algorithms are used to detect proper posture and proper movement~\cite{huang2011piano} (which we have also seen for other instruments such as violin~\cite{van2010musicjacket} and guitar~\cite{markyfrets2021, shinreal2018}), (ii) augmented visualizations have been overlaid to help learners with timing and which keys to press~\cite{weing2013piano}; (iii) gamified, practice and other modules have been introduced to motivate learners~\cite{rogers2014piano} and (iv) other interactions have been introduced to encourage users to improvise~\cite{karolus2020hit, deja2021encouraging, deja2021adaptive}. Whether these innovations have been very effective for their specific purpose warrants further investigation. 

What is known is that the piano has always been designed and created for the user taking various roles. 
A novice \textit{(learner)} learns the piano with the guidance of a mentor \textit{(teacher)}. A pianist performs in front of an audience -- \textit{(performer)}. In the midst of such performance, a pianist may improvise -- \textit{(improviser)}. A person composes a song and arranges it with the use of the piano -- \textit{(composer)}. While taking on one of these roles, users may have their own styles and systems to successfully accomplish their task.

As users typically experience pain points and problems when doing these tasks (teaching, learning, performing, improvising, composing)~\cite{chan2019applying}, technological innovations were introduced to potentially help them. However, newer innovations introduce newer affordances and issues with its use~\cite{dede1996evolution}, which we argue is a never-ending problem with technology. Additionally, not many features introduced by these innovations have been shipped as consumer-based features so far. To meet halfway, the existing use-cases in piano and the augmentations introduced to address the pain points users encounter along the way, we present our position towards a smart piano -- the human-centered piano of the future. We present different use-cases known (in the present) and emerging that we think may benefit from piano augmentation. We argue that such design may provide a more precise direction in the design and development of the piano of the future. 

\section{The Human-Centered Piano}
The piano is used somehow differently by users coming from various proficiency levels. A piano is a tool used to practice consistently by a novice or a beginner user. A more experienced pianist still trains with it but on a different cadence and intensity. They use the piano to practice for a performance (such as recitals or concerts) or to help them when composing new music. They may use the piano to record a performance, a melody or create unique arrangements. For both types of users, such tasks entail a lot of practice or trial-and-error activities. However, the depth in usage varies and is affected by other factors. Learning, though considered a personal activity, is usually a procedure that a mentor or teacher typically observes. Composing is either an activity done alone or in collaboration (if one has to play with others such as in a band, or a producer or sound engineer). During performances, the usage may involve certain constraints.

There is less pressure to make mistakes in front of a teacher during a teaching session than the amplified pressure a performer feels during a public performance. Improvising during a performance requires that the performer exhibits a solid mastery of music theory to produce musically sound and valid melodies. In terms of recording a composition, a musician should have practice their piece so the recording process becomes smooth and have lesser mistakes or trials. Overall, while all these involve the use of the piano, many factors are treated differently depending on each task. These include: (i) the ease of using the piano; (ii) the needed level of concentration during play; (iii) the confidence of the user during these activities; (iv) the cognitive load and other physiological factors experienced by the user.
As such, we need to introduce intelligent interventions that appeal to the different needs of these users. In the succeeding sections, we present how we envision some human-centered augmentations. 

\subsection{Learning}
There are several known methods for learning the piano such as Kodaly~\cite{howard1996kodaly} Dalcroze~\cite{mead1994dalcroze}, Orff~\cite{shamrock1997orff} and Suzuki~\cite{comeau2012playing}. These were implemented and have been augmented differently based on several learning frameworks and theories~\cite{bandura1977social, kolb2014experiential}. Generally, a piano learner spends a significant amount of time practicing, assessing their work and asking for support from the teacher. Self-reflection~\cite{zimmerman2009self} has been documented to help and motivate students as well.

Since learning the piano is usually a long and tedious process, the experience must be enriched to allow learners to be reflective, motivated and engaged. The ideal human-centered piano should enable learners to record their training sessions and reflect on them. Learners should be able to highlight specific segments in their practice recordings and seek support based on their performance. Constant recommendations regarding posture and body synchronization should be detected and provided to the learner, especially when a teacher is absent. The system should be able to capture and understand learners' current physiological state and use this information to manage their cognitive load. If possible, it should be captured in the most unobtrusive way possible without being compromised by noise. The captured data should be also used to manage the augmentations presented to the learner at a given time (e.g. bpm of overlaid piano roll visualizations, quantity and size of visualizations of keys that need pressing). 

While several of these features may have appeared in some prototypes published through the years, an ideal human-centered piano combines these and tackles the piano learning problem similar to that of learning a language or riding bike. Learners learns better the more they use the piano and build on the over-all experience. 

\subsection{Composing}
Music composition is summed up in three main activities: ideation, sketching and revision~\cite{bennett1976process}. When composing for the piano, a musician usually comes up with some notes and chords in segments. They write these notes with a pen and then switch to playing these keys on the piano to hear how it would sound. They begin with an initial sketch and cumulatively build on it. During this process they would switch between write something on paper and then playing it in the piano. This process repeats until they are satisfied. Or, in some instances, they would start over again. Mechanically, this process involves shifting from one task to another during the process -- from holding a writing tool for composing to pressing the piano keys and turning back. 

Based on the findings from~\cite{chan2019applying}, this usually leads to a loss of momentum or the musician's equivalent of a writer's block, especially for novices. In such scenarios, it would be ideal to have an AI agent that aids the musician in composing their piece without overtaking their tastes and appealing to their styles. This has been seen in some works~\cite{numao2002constructive, legaspi2007music, deja2018building} on emotion-based music systems as well. AI agents should not replace musicians but rather provide them with an augmented music composition process that makes them better in their work. A self-learning empathy framework can be incorporated in an human-centered piano to model, understand and then recommend musical hints to the composer (in the form of new keys, chords, motifs or segments). This can be achieved by integrating emotion detection techniques and utilizing existing computer-aided composition approaches. A human-centered piano should be equipped with sensors and other input modalities that would allow it to model, understand and build recommendations that would be personalized for a composer's styles and tastes. 

\subsection{Improvising}
A smart piano can have an AI agent that can collaborate with the user during composition and the actual performances. This is most important when a piano user improvises during a performance. Piano improvisation can often be regarded as a more advanced form of composition since the process takes place in play-time (playing the piano in real-time and not during practice). Improvisers perform not only the traditional melody defined in a sheet notation but use additional chords and/or melody that they think would make the piece more enjoyable or sound aesthetically different.

Based on music theory, certain chords and progressions go along together to create multiple or more harmonious melodies. Some combinations are not musically sound but still aurally acceptable (if assessed with quality of hearing output alone). This makes improvisation difficult to assess~\cite{deja2021encouraging} and augment. Like in composition, a human-centered piano should be a companion that the piano user can jam and ``collaborate'' with during a performance. This can be achieved by a piano that accepts inputs of different forms and modalities (such as physiological signals, spatiotemporal movement data, etc). For instance, it can be a gesture-based agent~\cite{karolus2020hit} that understands the \textit{free flowing movement of the arms of the piano user} much like dancing while performing and generating a sub melody in the process. It can be adaptive and dynamic that understands the individual taste of the user and then produce an impro piano roll helping the user to \textit{jazz-ify} the piece being played. It can also be an intelligent agent that lets the piano performer freely play their improvisation melody, records it and later on generates a \textit{reverse-piano roll} that the performer can save (and then review). This process would allow performers to evaluate and assess if their improvisation is musically sound based on music theory and rules. Additionally, the performers would broaden their musical vocabulary while getting intelligent feedback on their improvisation. 

\section{Discussion} 
We based the above presented visions of the human-centered piano with the ulterior goal of \textit{elevating the piano} from a tool to a companion. From the initial design of the piano, we have seen how different it is now. We introduce newer ways of interacting with the piano as we transition from simply producing a sound to recreating an existing masterpiece into a remixed symphony. We believe that the features we presented may represent that of a Human-Centered piano, however this is subject to further exploration and validation with actual users. We also posit that some benefits and disadvantages go with it as well. 

Experts and scholars argue that one essential component in piano playing is not losing its human touch~\cite{schubert2017algorithms}. As most compositions and pieces come from authentic human emotion, expressions and experiences, it is difficult to ignore that artificially generated motifs sequences may miss this vital component. This is why the human-centered piano must attempt to become a companion of the performer or composer and not entirely replace them. 

It will also be equally challenging to ensure that a user's unique identity (similar to how a piece can be uniquely classified as a work by Bach or Chopin) is not lost and merged with the shared repository of recommendations when used on a massive scale. For example, two talented performers must be able to use the human-centered piano and improve their musical vocabulary and at the same time have their work and performance easily distinguishable from each other.

Automatically-generated and emotion-inspired hints presented may have certain benefits as well. Piano users either spend less time in a musician's block when they receive tips or may be overwhelmed by the number of ideas. The proper timing and quantity of these hints must not overwhelm the user, and can be discovered through adequate user tests.

In the design of an human-centered piano, we factored in several modalities and parameters such as physiological data, motion data, personal data describing tastes and preferences. Security and privacy must be strictly-upheld in all these conditions. As we factor in more considerations in the design of this piano of the future, we introduce more benefits and issues with them. We posit that the piano of the future must always consider a holistic approach of the user: the learner, teacher, composer, performer and collaborator. 

\section{Summary}
We revisited the brief history of pianos and piano augmentation in this work. We drew inspiration for features of what would be an ideal human-centered piano. We looked at different activities, and use-cases musicians do with the piano and the augmentations that have been introduced with it. These activities and use-cases looked at piano users from different levels, namely students (as beginners), composers (as experienced users), and those who improvise (as even more advanced). We analyzed these critical points and identified how current or emerging technologies could be further developed towards a long-lasting design. Finally, we argue that a human-centered piano, designed towards the known use-cases at hand, would be an appropriate argument for the intelligent piano of the future. 

\bibliographystyle{ACM-Reference-Format}
\bibliography{main}


\end{document}